\documentclass[12pt]{article}
\usepackage{epsfig,psfrag}
\begin{document}

\begin{titlepage}
\begin{flushright}
NSF-ITP-02-36\\
ITEP-TH-17/02\\
\end{flushright}

\begin{center}
{\Large $ $ \\ $ $ \\
Operators with large R charge in $N=4$ Yang-Mills
theory.}\\
\bigskip
\bigskip
\bigskip
David~J.~Gross$\;^{\flat}$, 
Andrei~Mikhailov$\;^{\flat}$\footnote{On leave from 
the Institute of Theoretical and 
Experimental Physics, 117259, Bol. Cheremushkinskaya,
25, 
Moscow, Russia.}, Radu~Roiban$\;^{\sharp}$\\
\bigskip\bigskip\bigskip
$^{\flat}\;$ Institute for Theoretical Physics,\\
University of California, Santa Barbara, CA 93106\\
\bigskip
$^{\sharp}\;$ Physics Department,\\
University of California, Santa Barbara, CA 93106\\
\vskip 1cm
E-mail: $^{\flat}\;$gross@itp.ucsb.edu,
andrei@itp.ucsb.edu,
 $\;\;^{\sharp}\;$radu@vulcan.physics.ucsb.edu
\end{center}
\vskip 1cm
\begin{abstract}
It has been recently proposed that  string theory in 
the 
background of a  plane wave corresponds to a certain
subsector of the $N=4$ supersymmetric Yang-Mills
theory.
This correspondence follows as a limit of the AdS/CFT
duality.
As a particular case of the AdS/CFT correspondence, it
is a priori
a strong/weak coupling duality. However, the
predictions for
the anomalous dimensions which follow from  this
particular limit
are analytic functions of the 't Hooft coupling
constant $\lambda$ and  
have a  well defined expansion in the weak coupling
regime. 
This allows one to conjecture that the correspondence
between the strings on the plane wave background and
the
Yang-Mills theory works   at the level of perturbative
expansions.

In our paper we perform perturbative computations in
the Yang-Mills
theory that confirm this conjecture. We calculate the
anomalous
dimension of the operator corresponding to the
elementary
string excitation. We verify at the two loop level
that the
anomalous dimension has a finite limit when the R
charge
$J\to \infty$ keeping $\lambda/J^2$ finite. We
conjecture
that this is true at higher orders of   perturbation
theory. 
We show, by
summing an infinite subset of Feynman diagrams,  
under the
above assumption,  that the anomalous dimensions
arising from the
Yang-Mills perturbation theory are in agreement with
the anomalous
dimensions following from the string worldsheet
sigma-model.  
\end{abstract}
\end{titlepage}

\section{Introduction.}
\subsection{A new gauge fields $\leftrightarrow$
strings 
correspondence.}
The nature of the correspondence between gauge
theories and string
theory is one of the longstanding problems in modern 
theoretical physics. Significant progress was achieved in 
\cite{Maldacena,GKP,Witten} where the AdS/CFT
correspondence
was proposed. The AdS/CFT correspondence relates the
weak coupling limit of the string theory to the strong
coupling
limit of gauge theory, and  vice versa. On one hand,
this correspondence  is
useful because it teaches us about the strong coupling
behavior of   gauge 
theories (and probably the strong coupling limit of  
string theory). 
But on the other hand, it is often difficult to fully
exploit the duality. 
It is hard to quantize the superstring theory in
$AdS_5\times S^5$,
and therefore the calculations on the AdS side usually
do not go 
beyond the low energy supergravity approximation.
Also, it is hard to 
find independent confirmation of the correspondence
beyond
the agreement of those amplitudes that are protected
by
  supersymmetry.

An interesting proposal was made in \cite{BMN},
relating a
particular sector of the gauge theory to string theory
	in a plane wave
background. This can be considered a particular case
of the AdS/CFT
correspondence, because a plane wave is a limit of
AdS. 
The weakly coupled string theory is still mapped to
the
strongly coupled Yang-Mills in a sense that the
't~Hooft coupling
constant is large. However it turns out that in many
calculations
the effective coupling constant of the Yang-Mills
theory 
is not the 't~Hooft coupling $\lambda= g_{YM}^2N$ but
rather the product
$\phi^2\lambda$ where $\phi$ is a small number. This
allows  
perturbative computations to be extended to the large
$\lambda$
region. When $\phi^2\lambda$ is small, the
correspondence of
\cite{BMN} maps some perturbative computations on the
Yang-Mills
side to   superstring perturbation theory. Moreover,
it turns out
that the string worldsheet sigma-model is exactly
solvable in
the plane wave background
\cite{Metsaev},\cite{RUTS}-\cite{KIPI}. 
This allowed the authors of \cite{BMN} 
to construct the explicit map 
between the string states and   gauge invariant
operators in
the Yang-Mills theory.  

The proposal of \cite{BMN} was subsequently
extended to other gauge/gravity dualities. Backgrounds
with 
minimal supersymmetry
were first discussed in \cite{IKM}-\cite{PASO}.
The Penrose limits of orbifolds of $AdS_5\times S^5$
were 
discussed in \cite{ASJ}-\cite{MICH} and the operators
dual 
to elementary string excitations were also
constructed.
Other spaces, arising from brane 
intersections \cite{LEPA}, \cite{LUPO}, spaces
describing 
gauge theory RG flows \cite{GUNS} and the
Randall-Sundrum 
scenario \cite{GUVE} were analyzed. String
couplings to D-branes were studied in \cite{IMAM}
confirming 
the conjectured correspondence between string modes
with high 
R-charge and superYang-Mills operators. Interactions
of strings in a
plane wave background were discussed in \cite{SPVO}
where propagators
and closed string vertices were constructed.
First steps towards the derivation of the string interactions
from the field theory were taken in \cite{KPSS}.
The interesting question of holography 
in plane wave background was attacked from various
perspectives
in \cite{DAGR}, \cite{LEOR}.

\subsection{Anomalous dimensions from quantum
mechanics.}
Once the correspondence is established, the first
nontrivial check
is whether the conformal dimension of the operator is
in agreement
with the mass of the corresponding superstring state
\cite{Witten}.
The authors of \cite{BMN} invented a beautiful trick
which allowed them to compute the anomalous dimension
of certain
non-BPS operator exactly in perturbation theory. They
have found 
complete agreement with the string theory calculation.

But their argument relied on a certain assumption
about the 
behavior of   anomalous dimensions in the Yang-Mills
theory 
in the strong coupling
limit, which itself follows from the AdS/CFT
correspondence and
has not been independently verified. 
We will now review the arguments of \cite{BMN}.

The crucial ingredient in the BMN proposal is the
observation that 
single 
trace operators with parametrically large R-charge can
be put in one to 
one correspondence with {\it physical} string states.
The construction 
of 
the plane wave limit on $AdS_5\times S^5$ isolates an
$SO(2)$ subgroup 
of the $SU(4)$ R-symmetry group. Furthermore, the
limit keeps a subset of 
the YM operators--- those with $SO(2)$ charge of the
order of the square 
of the $AdS$ radius, i.e. of the order $N^{1/2}$. The
BPS bound then 
implies that the operators kept in the limit are those
with conformal 
dimension $\Delta$ of the order $N^{1/2}$. It is then
natural to 
collect together operators with the same difference
$\Delta-J$.

As a limit of $AdS_5\times S^5$, the plane wave
background is supported 
by a (null) RR flux. As usual, the quantization of the
NSR string in 
such a background is problematic. However, one can use
the GS model 
constructed for $AdS_5\times S^5$ as a starting point.
Taking the plane 
wave limit here leads to substantial simplifications.
Choosing 
light-cone
gauge for $\kappa$-symmetry leads to a quadratic
action which is easily 
quantizable. This approach was pursued in
\cite{Metsaev}, where 
the spectrum was constructed, with the result that
acting with a
level 
$n$ creation operator on some state adds
\begin{equation}\label{StringPrediction}
\delta m^2_n=\sqrt{\mu^2+\frac{n^2}{(\alpha'p_+)^2}}
\end{equation}
to the mass of the ground state. 
According to the usual AdS/CFT philosophy one should
match this with 
the anomalous dimension of the operator dual to this
state. 
Written in terms of the parameters of the Yang-Mills
theory 
the anomalous dimension predicted by the string theory
formula
(\ref{StringPrediction}) is:
\begin{equation}\label{YangMills}
(\Delta-J)_n=\sqrt{1+{4\pi gNn^2\over J^2}}
\end{equation}
The authors of \cite{BMN} suggested an elegant way to
compute 
the anomalous dimension of nearly BPS operators in
$N=4$ theory
which gives the answer in agreement with
(\ref{YangMills}).
Anomalous dimensions of operators are the same as the
energies of
the corresponding states in the field theory on ${\bf
R}\times S^3$.
The proposal of \cite{BMN} is to reduce
the field theory to   quantum mechanics by taking into
account only 
zero modes on $S^3$. Then one computes the corrections
to the ground state energy in  quantum mechanical
perturbation 
theory. 
How this works can be understood, for example, from
the second order 
correction to the energy: 
\begin{equation}\label{QMsum}
\sum_i\frac{\langle E_j|E_i'\rangle\langle
E_i'|E_j\rangle}{E_j-E_i'}
\end{equation}
In principle, this correction receives contributions
from all  
states
$|E_i'\rangle$ orthogonal to $|E_i\rangle$. 
However, the contribution is suppressed by the inverse
of the
energy of the state. The standard AdS/CFT
correspondence tells us that 
 states which involve   nonzero modes on $S^3$ have a
very large 
energy
in the strong coupling limit.  
Therefore the contribution of these states to 
(\ref{QMsum}) is very 
small
and we can neglect them. Neglecting the states created
by  the nonzero 
modes
of the Yang-Mills fields on $S^3$ should be equivalent
to dimensionally
reducing the Yang-Mills theory down to quantum
mechanics.
This is summarized in the following cartoon
where the box stands for the corrections that remove
the massive 
spectrum resulting from the reduction of the YM theory
on $S^3$:
%While this argument may be correct 
%and certainly leads to agreement with the string
%computation, it is 
%not clear 
%why one is allowed to use in the second order of this
%perturbative 
%expansion 
%the result of the resumation of the diagrams leading
%to the divergence 
%of
%the energy of the massive states. In particular,
%since there are no 
%infraread divergences in the correlation
%functionsleading to 
%corrections
%to the energy of the states, it is not {\it a priori}
%clear why one is 
%allowed to resum diagrams in the first place. This is
%summarized in 
%the 
%following cartoon:
\begin{figure}[ht]
\begin{center}
\psfrag{a}{Strong coupling}
\epsfig{file=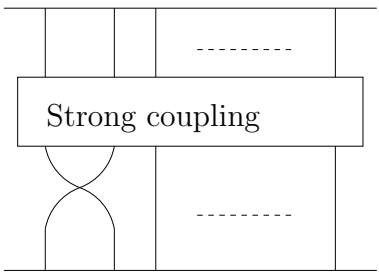,width=5cm}
\caption{``Dynamical removal'' of massless states.}
\end{center}
\end{figure}

\subsection{Our paper.}

The correpondence between the gauge theory and the plane wave background was 
derived from the AdS/CFT correspondence using the supergravity approximation
on the AdS side.  
When the 't Hooft coupling parameter is small the 
supergravity approximation breaks down. 
However one can conjecture that the BMN state is still approximated by the
string moving in the pp-wave background,
even though this was derived in the supergravity approximation.
This would imply that the formula (\ref{YangMills}) is reproduced by the
standard 't Hooft perturbative expansion.

The authors of \cite{BMN} also used the perturbative expansion
to derive (\ref{YangMills}) on the CFT side, but that was not the
Yang-Mills perturbative expansion. In their argument it was essential
that one first makes the dimensional reduction from the $1+3$ dimensional
Yang-Mills theory to the $1+0$ dimensional quantum mechanics.  
In fact it turns out that after the dimensional reduction the
perturbative expansion is justified even when $g_{YM}^2N$ is large.
One finds that the small parameter governing the series 
expansion is in fact not $\lambda$ but rather ${n^2\over J^2}\lambda$. 
This is due to the locality of the quantum mechanical Hamiltonian. 
The BMN Hamiltonian
\begin{equation}
H_{BMN}=\sum_j b_j^{\dagger} b_j + {gN\over
2\pi}(b_j+b_j^{\dagger}-
b_{j+1}-b_{j+1}^{\dagger})^2
\end{equation}
contains  a maximum of two derivatives after taking the continuum limit. 
And $gN$ appears only multiplying the square of the lattice spacing. 
(This is the reason why the continuum limit exists.) This simple form 
of the Hamiltonian is in turn due to the simplicity of the Feynman 
diagramms with the non-zero modes projected out.
 
In principle one can imagine that taking into account the massive 
states leads to some more complicated Hamiltonian for which
the continuum limit does not exist. For example, a perturbation of 
the form
\begin{equation}\label{BadPerturbation}
\Delta H= \sum_j \left({gN\over 2\pi}\right)^2
\left[(b_j+b_j^{\dagger})^2-(b_{j+1}+b_{j+1}^{\dagger})^2\right]^2
\end{equation}
would become infinitely strong in the BMN regime, 
invalidating the perturbation theory. 

Suppose that we work at weak 't~Hooft coupling, and write down
order by order in perturbation theory the effective Hamiltonian
governing the renormalization of the composite operator.
Is it true that this effective Hamiltonian will have a nice continuum
limit, or will it contain terms like (\ref{BadPerturbation})?

In our paper we verify that in the lowest nontrivial
order of perturbation 
theory (two loops)   the renormalization does indeed
have 
a continuum limit. We   find that the anomalous
dimension
of the operator dual to the string state is indeed a
function
of $\lambda$ and $n/J$ in the combination $\lambda
n^2/J^2$.

We then assume that this is true to higher orders of  
perturbation
theory. By summing up an infinite subset of Feynman
diagrams we show 
that, under this assumption, the anomalous dimension
of the
operator is indeed given by (\ref{YangMills}). 

Our results provide evidence for the conjecture that
the correspondence between gauge fields and strings
proposed
in \cite{BMN} works in perturbation theory.

We should stress that there have been a number of
interesting papers
on perturbation theory for $N=4$ Yang-Mills in the
context of
the AdS/CFT correspondence. 
In particular, a remarkable computation of the 
circular Wilson loops to all orders in perturbation
theory was performed in \cite{circular}. 
The absence of the perturbative corrections in the order $g_{YM}^2$ 
to the two and three point functions of the chiral primary operators 
was demonstrated in \cite{DHFS}. Two point functions of 
chiral primary operators were computed to the order $g_{YM}^4$ 
in \cite{PSAZ}. 
A perturbative computation of the correlation
functions of the BPS operators at two loop order was
performed in superspace in \cite{StonyBrook}. 
The superspace approach to the computation of the
anomalous dimension of the composite operators was developed in
\cite{PenatiSantambrogio}.  

\section{Anomalous dimension from  Yang-Mills
perturbation 
theory.\label{generalandim}}

\subsection{General facts about the anomalous
dimension.}

The maximally supersymmetric Yang-Mills theory
contains six 
real scalars which we denote $\phi^1,\ldots,\phi^6$.
Let us concentrate 
on a $U(1)$ subgroup of the $SU(4)$ R-symmetry group
which rotates 
$\phi^5,\phi^6$ and leaves the other four scalars
invariant. It is 
natural to
construct $Z=\phi^5+i\phi^6$, which has unit charge
with respect to 
this subgroup.

\begin{figure}[ht]
\begin{center}
\epsfig{file=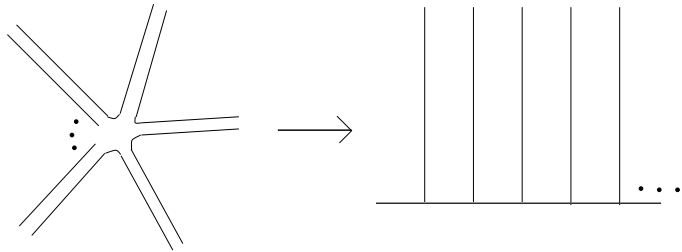,width=9cm}
\caption{Simplified diagramatic notation 
for operators.\label{operators}}
\end{center}
\end{figure}

It was argued in \cite{BMN} that the
Penrose limit of the AdS geometry corresponds, on 
the gauge theory side, to focusing on
the set of operators with large R-charge.
More precisely, the $U(1)$-component of the $R$ charge
should be
very large while the other components should be of 
order one. From
this perspective the ground state of the string can be
put in 
correspondence with the BPS operator $\mbox{tr}\;Z^J$,
$J\gg 1$. 
Other massless modes as well as excited string modes 
correspond to inserting $\phi^a,\,a=1,\dots,4$ and 
$\oint_{S^3}n^a Z(n)$ into the ``string'' of $Z$'s in
a very specific 
way.
It is further assumed that the number of such
insertions is small.
One of the many possible operators obtained in this
way is
\begin{equation}
{\cal O}_m=\mbox{tr}\;\phi Z^m \phi Z^{J-m}
\end{equation}
where we have inserted two $\phi$ fields. 
We will use a schematic notation for such operators
which is shown in 
figure \ref{operators}. An insertion of a composite
operator will 
be denoted by a horizontal line. Its intersection
points with other 
lines are understood as being at the same space-time
point. 

We will be interested in computing the anomalous
dimensions of 
operators dual to excited string states. Starting from
the requirement 
that these operators have definite anomalous
dimensions and from the 
assumption \cite{BMN} that they are linear
combinations 
of the operators described above, we will recover the
full set of
operators conjectured in \cite{BMN}.

An   aspect of renormalization of composite operators
that often 
  appears in   ordinary quantum field theories
is operator mixing: the divergences in 1PI diagrams
with one insertion 
of a composite operator generally contain divergences
proportional to 
other composite operators. Thus, all composite
operators must be
renormalized at the same time, order by order in
perturbation theory.
Furthermore, one has to take into account the usual
field and coupling 
constant renormalization. This leads to  the following

renormalization:
\begin{equation}
{\cal
O}_i^{bare}(\Phi^{bare},g^{bare})=\sum_k\,Z_i{}^k
{\cal O}_k^{ren}(Z_3^\Phi\Phi^{ren}, {Z_1\over 
Z^\Phi_3{}^{3/2}}g^{ren})
\label{renormalization}
\end{equation}
where $\Phi$ denotes a generic field, we assume   the
existence of 
a cubic $\Phi$ interaction of strength $g$ and $Z_3$
and $Z_1$ are the 
usual
wave function and coupling constant $Z$-factors.

\begin{figure}[ht]
\begin{center}
\epsfig{file=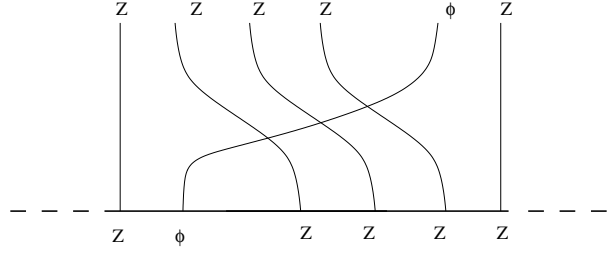,width=8cm}
\caption{Contribution to $c_{3,3}$.\label{c33}}
\end{center}
\end{figure}

We will proceed by considering the operators ${\cal
O}_m$ 
introduced above, but the arguments generalize
immediately
to more complicated ones. To cancel divergencies in
the 
proper graphs with one ${\cal O}_m$ insertion we need
to add as
counterterms local operators with the same engineering
dimension 
and R-charge as ${\cal O}_m$. However, the only
operators with 
R-charge  $(J,2,0)$ and dimension $J+2$ are of the
type ${\cal O}_n$ 
for some $n$. This implies that the counterterms
needed to cancel
divergences in a 1PI graph with one insertion of
${\cal O}_m$ 
will be   linear combinations of ${\cal O}_n$. It is
not hard to see 
that, for $m$, $J$ and $J-m$ sufficiently large,
planar diagrams are 
invariant  under $m\rightarrow m+1$. Thus, all
counterterms should 
have this symmetry. This observation implies that the
only operator
which is multiplicatively renormalized is, up to
overall normalization,
\begin{equation}\label{Wave}
{\cal O}(k)=\sum_m e^{2\pi imk/J}\mbox{tr}\;\phi Z^m
\phi Z^{J-m}
\end{equation}
for which equation (\ref{renormalization}) becomes
\begin{equation}
\label{renop}
{\cal O}(k)^{bare}=
Z_{\cal O}Z_3^\phi(Z_3^Z)^{J/2}{\cal O}(k)^{ren}\equiv
Z(\lambda,\epsilon){\cal O}(k)^{ren}~~.
\end{equation}
Standard manipulations now imply that the anomalous
dimension 
${\cal O}(k)$ is:
\begin{equation}
2c(\lambda)=\epsilon{d\log Z\over d\log\lambda}~~.
\end{equation}
For later convenience let us introduce the notation
$$
\varphi={2\pi \over J}
$$
In the expansion 
$$
c(\lambda)=c_1(k)\lambda+c_2(k)\lambda^2+\ldots
$$
the coefficients $c_1,c_2,\ldots$ are functions of
$e^{ik\varphi}$:
\begin{equation}
c_j(k)=\sum\limits_{n=-j}^j c_{j,n}e^{ink\varphi}
\end{equation}
The coefficient $c_{j,j}$ can be easily computed in
perturbation
theory for arbitrary $j$, because only one diagram
contributes to it. 
For 
example, the contribution to $c_{3,3}$ is given by the
graph in figure 
\ref{c33}.
%\begin{flushleft}
%\epsffile{ladder.eps}
%\end{flushleft}
%(This picture shows the contribution to $c_{3,3}$.)
We will compute the contribution of these diagrams in
Section 5.

\subsection{A prediction from the dual string theory.}

In equation (\ref{StringPrediction}) the anomalous
dimension depends
on $\varphi$ and the 't~Hooft coupling in the
combination $\varphi^2 
g_{YM}^2N$.
If (\ref{StringPrediction}) is satisfied order by
order in  
Yang-Mills perturbation theory this would mean that
\begin{equation}\label{Conjecture}
c_j(k)=c_{j,j}(e^{ik\varphi}+e^{-ik\varphi}-2)^j
\;\;\;\;\;\;\mbox{(conjecture)}
\end{equation}
In other words
\begin{equation}
\label{coefconj}
c_{j,n}=(-1)^{j-n} c_{j,j}{(2j)!\over (j-n)!(j+n)!}
\;\;\;\;\;\;\mbox{(conjecture)}
\end{equation}
In Section \ref{2loops} we will compute the
coefficients $c_{2,1}$ and 
$c_{2,2}$ and show that the conjecture
(\ref{coefconj}) holds at the
two loop level.

\section{The anomalous dimension at one loop.}

\subsection{The Feynman rules.}
\begin{figure}[ht]
\begin{center}
\epsfig{file=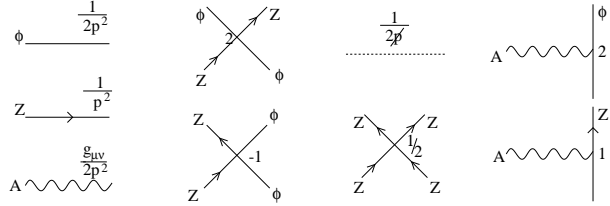,width=8cm}
\caption{Feynman rules relevant for our
computations.\label{rules}}
\end{center}
\end{figure}

The lagrangian of the $N=4$ Yang-Mills theory can be
derived in a 
number of ways. In the following we will use the form
that arises
by dimensionally-reducing ten-dimensional
super-Yang-Mills
theory on a six-dimensional torus:
\begin{equation}
\begin{array}{l}
L={1\over g_{YM}^2}\int\mbox{tr}\left\{
-(\partial_{\mu}A_{\nu})^2+
(\partial_{\mu}\phi^I)^2+\bar{\psi}\hat{\partial}\psi\right.
+\\[5pt]+
\left.
2
A_{\mu}(\phi^I\stackrel{\leftrightarrow}{\partial}\phi^I)+
\bar{\psi}[\hat{A},\psi]+\bar{\psi}\Gamma^I[\phi^I,\psi]+
{1\over 2}[\phi^I,\phi^J]^2+{1\over
2}[A_{\mu},A_{\nu}]^2\right\}
\end{array}
\end{equation}
Rewriting this lagrangian in terms of
$Z=\phi^5+i\phi^6$, 
introduced in the previous section, we find that the
two-point 
functions of fields $\phi$ are
$\langle\phi\phi\rangle={1\over 2p^2}$
and the two-point function of $Z$ is $1\over p^2$.
The set of relevant Feynman rules is summarized in
figure \ref{rules}.
%\begin{flushleft}
%\epsffile{rules.eps}
%\end{flushleft}
We choose to work in Feynman gauge
due to the simplicity of the gauge boson propagator.
One can use a 
general 
renormalizable gauge as the result is   independent of
the choice
of gauge.

\subsection{Our notation for integrals.}
Perturbation theory is an expansion in 
${g^2\over (4\pi)^2}$ where the denominator arises
from 
integrals over loop momenta. To simplify notation  we
will include this factor
 in the coupling constant from the outset. Thus, in
our notation
the loop integrals look like:
\begin{equation}
\int d^{4-2\epsilon}q\,{1\over q^2(q+p)^2}=
{\Gamma(\epsilon)\Gamma(1-\epsilon)^2\over\Gamma(2-2\epsilon)}
{1\over [p^2]^{\epsilon}}~~.
\end{equation}
We have further absorbed a factor of $(-4\pi\mu^2)$
into $p^2$ making 
it 
dimensionless. At higher loops other integrals become
useful and they 
will be introduced as we go along.

We are now ready to compute the one-loop contribution
to the anomalous 
dimensions of the operators introduced in \cite{BMN}
and described in 
detail in 
section \ref{generalandim}.

\subsection{The anomalous dimension at one loop.}
Before we proceed with the calculations we want to
discuss the role
of the ``dilute gas'' assumption which is appropriate
when  $J\gg 1$.

As discussed in section \ref{generalandim} we  
consider operators
with a small number of $\phi$-fields. 
In low orders of   perturbation theory it is easy to
see that the
diagrams responsible for the anomalous dimension of
${\cal O}$
involve essentially only the $\phi$ fields and a few
$Z$ fields
next to the $\phi$ insertion. 
This implies that, for small enough
number of loops, it is enough to study operators with
exactly one
$\phi$-field.
Such an operator vanishes due to the cyclicity of the
trace. 
Therefore,
to 
obtain a meaningful result we write the operator as
\begin{equation}
{\cal O}=\sum e^{il\varphi }  Z^l\phi Z^{J-l}~~.
\end{equation}
Even though this operator is not gauge invariant, it
is meaningful to 
discuss it because for a small enough number of loops
the required 
counterterms are proportional to ${\cal O}$. It can be
interpreted as a 
building block for operators with a larger number of
$\phi$ insertions. 

If $J$ is finite, then there is
the danger that at $J$ loops two $\phi$ fields arrive
next to each 
other 
invalidating the initial assumption of a  "dilute
gas". 
In fact, we do expect such ``contact'' terms to affect
the anomalous
dimension. For example, consider the operator with two
insertions
of $\phi$ symmetrized over the positions of $\phi$.
The ``bulk'' 
contribution
to the anomalous dimension of such an operator is
zero. (Indeed, the
bulk contribution is just twice the anomalous
dimension of 
$\mbox{tr}\;\phi Z^J$ which vanishes because 
$\mbox{tr}\;\phi Z^J$
is BPS.) But the operator with   two insertions of
$\phi$ is not
BPS, therefore it should acquire an  anomalous
dimension. 
We expect that this anomalous dimension arises
precisely from  
diagramms with the two $\phi$'s appearing next to each
other.

Since we  e neglect  the contact terms, we can only
say that our
result for the anomalous dimension is correct up to,
roughly, the $J$-th 
order
of   perturbation theory. 

Let us now proceed with the one loop computation.
There are two classes of diagrams, with the exchange
of $Z$ and $\phi$ 
and without the exchange. We will start by computing
the former.
There is exactly one diagram which mixes the original
operator  
with operators in which $\phi$ is moved one site to
the left or to 
the right.
\begin{figure}[ht]
\begin{center}
\epsfig{file=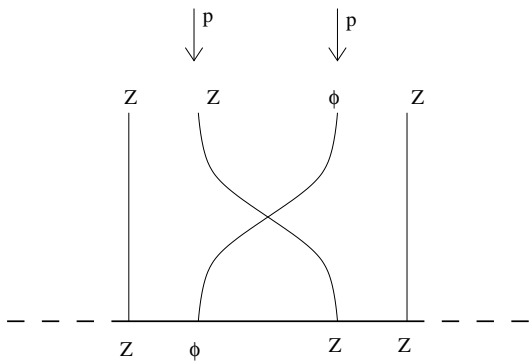,width=7cm}
\caption{One loop diagram with the exchange of $Z$ and
$\phi$.}
\end{center}
\end{figure}
%\begin{center}
%\leavevmode
%\epsffile{oneloop.eps}
%\end{center}
The amplitude is:
\begin{eqnarray}
&&
I_1(2p)=\lambda e^{i\varphi}\int
{d^{4-2\epsilon}q\over q^2 (q+2p)^2}
=\lambda e^{i\varphi} 
{\Gamma(\epsilon)\Gamma(1-\epsilon)^2\over
\Gamma(2-2\epsilon)}\left({1\over
4p^2}\right)^{\epsilon}=
\nonumber\\[5pt]&&=
\lambda e^{i\varphi}\left[{1\over\epsilon}
+2-C+\log{1\over 4 p^2}\right]
\label{ex}
\end{eqnarray}
We now turn to diagrams not exchanging $Z$ and $\phi$.
These can be 
split again
in two subclasses. 
The first subclass involves interactions of $\phi$ and
$Z$ while 
the second one contains only interactions among
$Z$-fields.

There are 
two diagrams in the first subclass (Fig. 6). The first
one, like the 
diagram discussed above, arises from the 
four-scalar interaction. The second one
has a gluon exchange.
\begin{figure}[ht]
\begin{center}
\epsfig{file=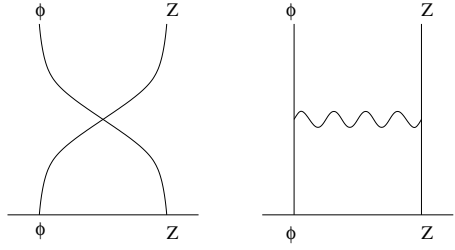,width=6cm}
\caption{One loop diagram not exchanging $Z$ and
$\phi$.}
\end{center}
\end{figure}
%\begin{center}
%\epsffile{noexchange.eps}
%\end{center}
To find the anomalous dimension, we need only the
divergent part.
The divergent part of the scalar diagram is 
minus the divergent part of the diagram with the
vector exchange
and therefore the sum of these two diagrams is finite.

As for the diagrams with   two $Z$ lines there are
again two of 
them;
the diagram with a four scalar interaction and the
diagram with  a
gluon exchange. They both appear with the same sign,
and the divergent 
part
is
\begin{equation}
\label{Znoex}
I_Z(2p)={\lambda\over\epsilon}+(finite)
\end{equation}

\begin{figure}[ht]
\begin{center}
\epsfig{file=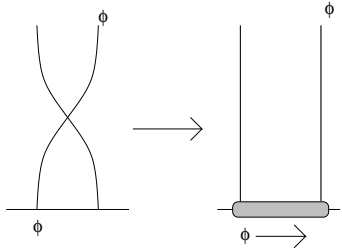,width=4.5cm}
\caption{Counterterm with exchange of $\phi$ and
$Z$.\label{exct}}
\end{center}
\end{figure}

To summarize, we need the following counterterms: 

$\bullet$ one counterterm in which $\phi$ is moved one
site to the 
right (figure \ref{exct}).
%\begin{flushleft}
%\epsffile{ct1.eps}
%\end{flushleft}
Its value is equal to $-{1\over\epsilon}e^{i\varphi}$.

$\bullet$  one counterterm in which
$\phi$ is moved one site to the left, which is the
complex conjugate of 
the 
above one.  

\begin{figure}[ht]
\begin{center}
\epsfig{file=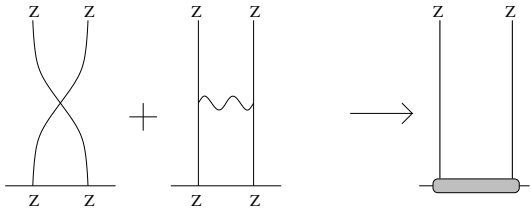,width=7cm}
\caption{Counterterm for $Z$ interaction.\label{Zct}}
\end{center}
\end{figure}

$\bullet$ $J-2$ counterterms with interactions between
two $Z$ lines 
in which $\phi$ keeps its original position.
They are due to interactions among $Z$-fields (figure
\ref{Zct}).
%\begin{flushleft}
%\epsffile{ctZ.eps}
%\end{flushleft}
Each such counterterm is equal to $-{1\over\epsilon}$.

Putting together 
equations (\ref{ex}) and (\ref{Znoex}) it is not hard
to see that
we will cancel the divergence in the diagrams
containing one insertion
of ${\cal O}$ if we define the renormalized $Z_{\cal
O}$ in 
equation (\ref{renop}) as:
\begin{equation}
\label{ZO}
Z_{\cal
O}=\left(1-{\lambda\over\epsilon}(e^{i\varphi}+e^{-i\varphi}
+(J-2))\right)
\end{equation}
The last step is finding the wave function
renormalization $Z_3^\phi$ 
and $Z_3^Z$. The full correction to the propagator
will turn out to be 
useful in the two loop computation so we will write it
down in detail.

\subsection{One loop corrections to the propagator and
the anomalous dimension.}
Let us start with computing the one loop correction to
the propagator 
of
the scalar field ($\phi$ or $Z$). There are three
diagrams
\begin{equation}
{1\over p^2}+{\lambda\over p^4}\left[\int d^4k
{(k+2p)^2\over k^2(k+p)^2}+8\int d^4k {k(k+p)\over
k^2(k+p)^2}
-9\int {d^4k\over k^2}\right]
={1\over
p^2}-2\lambda\left[{1\over\epsilon}+2-C\right] 
{1\over (p^2)^{1+\epsilon}}
\end{equation}
where the first term in parentheses 
comes from a gauge boson loop, the second one
from a fermion loop while the third one from a scalar
tadpole. 
Therefore 
the renormalization of the propagator is:
\begin{equation}
\label{1loopscalarprop}
Z_3^\phi=Z_3^Z=1+2{\lambda\over \epsilon}
\end{equation}
Combining the analogue of the equation (\ref{renop})
for the operator 
considered here
(only one insertion of $\phi$) with equations
(\ref{ZO}) 
and (\ref{1loopscalarprop})
we find that 
\begin{equation}
Z(\lambda,\epsilon)=Z_{\phi}^{J/2}Z_{\cal O}=
1-{\lambda\over\epsilon}(e^{i\varphi}+e^{-i\varphi}-2)
\end{equation}
which implies that the one loop contribution to the
anomalous dimension
of the operator ${\cal O}$ is
\begin{equation}
c_1(\lambda)={\lambda\over\epsilon}(e^{i\varphi}+e^{-i\varphi}-2)\,
,
\end{equation}
in accord with equation (\ref{Conjecture}). The fact
that $c_1$ depends
on $\lambda$ only as $\lambda
(e^{i\varphi}+e^{-i\varphi}-2)$ is at one
loop level a consequence of supersymmetry.

We are now ready to discuss the two loop contribution
to the anomalous 
dimension.

\section{Two loops.\label{2loops}}
As discussed in section (\ref{generalandim}), the two
loop contribution 
to anomalous dimensions has a natural expansion in
terms 
of $e^{i\varphi}$ as
\begin{equation}
c_2=c_{2,0}+c_{2,1}(e^{i\varphi}+e^{-i\varphi})+
c_{2,2}(e^{2i\varphi}+e^{-2i\varphi})
\end{equation}
In the next subsection we will compute the last 
coefficient, $c_{2,2}$. We will 
continue by computing $c_{2,1}$ and finish by
extracting $c_{2,0}$ from 
the fact that, for $\varphi=0$, the operator ${\cal
O}$ is BPS and thus 
has vanishing anomalous dimension.

\subsection{Diagrams proportional to $e^{2i\varphi}$.}

\begin{figure}[ht]
\begin{center}
\epsfig{file=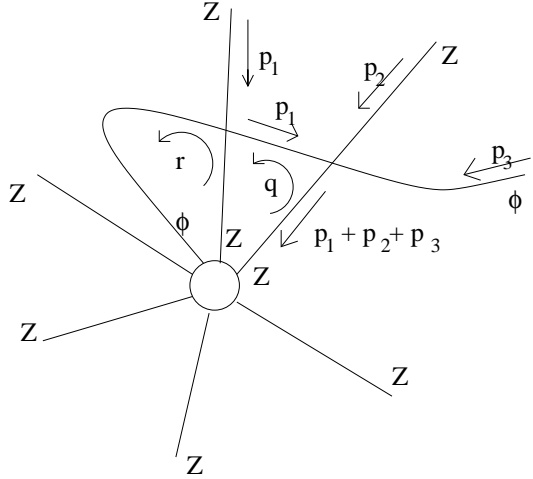,width=7cm}
\caption{The unique contribution to
$c_{2,2}$.\label{e2iphi}}
\end{center}
\end{figure}

As stated in section \ref{generalandim}, there is just
one diagram 
contributing to $c_{2,2}$. This diagram involves only
interactions of
scalar fields and is shown in figure \ref{e2iphi}.
%\begin{center}
%\leavevmode
%\epsffile{dia.eps}
%\end{center}

\noindent
The amplitude reads:
\begin{eqnarray}
I_2(p_1,p_2,p_3)=
\lambda^2 e^{2i\varphi} \int {d^{4-2\epsilon}r
d^{4-2\epsilon}q 
\over r^2(r+q)^2(q-p_1)^2(q-(p_1+p_2+p_3))^2}
=\nonumber\\[5pt]=
\lambda^2 e^{2i\varphi}
{\Gamma(\epsilon)\Gamma(1-\epsilon)^2
\over \Gamma(2-2\epsilon)}
\int {d^{4-2\epsilon}q 
\over [q^2]^{\epsilon}(q-p_1)^2(q-(p_1+p_2+p_3))^2}
\end{eqnarray}
To analyze the properties of this last integral, 
let us consider the generic expression
\begin{equation}
\int {d^{4-2\epsilon}q 
\over [q^2]^{\epsilon}(q-p)^2(q-p')^2}~~.
\end{equation}
This integral is logarithmically divergent, and up to 

terms proportional to $\epsilon$ it is a function of 
$p-p'$. Indeed,
\begin{eqnarray}
\int d^{4-2\epsilon}q \left({\partial\over\partial
p_{\mu}}
+{\partial\over\partial p'_{\mu}}\right)
{1 \over [q^2]^{\epsilon}(q-p)^2(q-p')^2}
=
-\int {d^{4-2\epsilon}q\over [q^2]^{\epsilon}} 
{\partial\over\partial q_{\mu}}
{1 \over (q-p)^2(q-p')^2}
=\nonumber\\[5pt]=
-\epsilon\int d^{4-2\epsilon}q
{2q^{\mu} \over [q^2]^{1+\epsilon}(q-p)^2(q-p')^2}
\end{eqnarray}
The integral is convergent, therefore this expression
is zero up
to  terms of order $\epsilon$. Such precision is
enough for the
computation of anomalous dimensions at this order, and
 therefore 
$I(p_1,p_2,p_3)$ is, for our purposes, a function of
$p_2+p_3$:
\begin{eqnarray}
I(p_1,p_2,p_3)=
\lambda^2 e^{2i\varphi}
{\Gamma(\epsilon)\Gamma(1-\epsilon)^2
\over \Gamma(2-2\epsilon)}
\int {d^{4-2\epsilon}q 
\over [q^2]^{\epsilon+1}(q-(p_2+p_3))^2}
=\nonumber\\[5pt]=
\lambda^2 e^{2i\varphi}{1\over 
[(p_2+p_3)^2]^{2\epsilon}}
{1\over\epsilon(1-2\epsilon)}{\Gamma(2\epsilon)\Gamma(1-\epsilon)^3\over
\Gamma(2-3\epsilon)}
\end{eqnarray}
Subtracting the counterterm in figure \ref{exct} we
arrive at a local 
divergence:
\begin{eqnarray}\label{ITwoSubtr}
I_2(p_1,p_2,p_3)-{\lambda e^{i\varphi}\over
\epsilon}I_1(p_2+p_3)=
\lambda^2e^{2i\varphi}\left[
-{1\over 2\epsilon^2}+{1\over
2\epsilon}+\mbox{finite}\right]~~.
\end{eqnarray}
One can easily extract from here the contribution to
$c_{2,2}$. We 
will, 
however, postpone this to the end of this
section when we will find
the full two-loop result.

\subsection{Diagrams with $e^{i\varphi}$.}
The set of diagrams leading to a shift in the position
of $\phi$ by
one site can be naturally decomposed in three disjoint
sets: diagrams 
with only two interacting legs, diagrams with three
interacting legs 
and 
diagrams with disconnected one-loop graphs.
Once the counterterms are included, the three sets of
diagrams lead 
only
to local divergences. This is the case since there are
no one-loop
counterterm  graphs mixing any two of the three sets
of diagrams.

We should take into account the contribution of the
fermions.
In computing the fermionic loops we will use the
dimensional 
regularization 
via the dimensional reduction which was first
suggested in 
\cite{Siegel}.
It uses the four-dimensional algebra of gamma-matrices
and 
four-dimensional
tensor algebra, but the momenta are taken to be 
$(4-2\epsilon)$-dimensional.
This regularization is consistent at low orders of
perturbation theory 
as
long as antisymmetric Levi-Civita tensors are not 
present,
and manifestly preserves supersymmetry. (See the
discussion in 
\cite{AvdeevVladimirov}.) 

\begin{figure}[ht]
\begin{center}
\epsfig{file=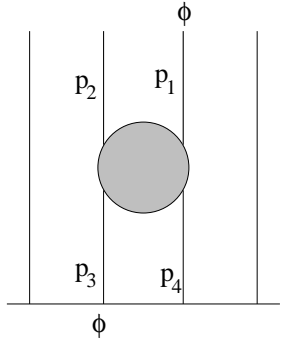,width=3.7cm}
\caption{Generic  diagram with ``two interacting
legs''.\label{2leg}}
\end{center}
\end{figure}

We will begin by computing the  diagrams with ``two
interacting legs'' 
shown in figure \ref{2leg}.
%\begin{flushleft}
%\epsffile{vertex.eps}
%\end{flushleft}
In this figure the dashed circle denotes the one loop
corrected 
interaction vertex of four scalars. 
We will start with computing this object. In this
computation we will 
use the
following convention about the external lines. Two
external lines 
carrying momenta $p_1,p_2$ are considered amputated. 
The other two external lines carrying momenta $p_3$
and $p_4$ are the
propagators of $\phi$ and $Z$ respectively. 

For further convenience let us introduce the notation:
\begin{equation}
[q_1,q_2,q_3,q_4]:= (q_1\cdot q_2)(q_3\cdot q_4)+
(q_1\cdot q_4)(q_2\cdot q_3)-(q_1\cdot q_3)(q_2\cdot
q_4)
\end{equation}

\noindent
The relevant 1PI diagrams are:
\begin{figure}[ht]
\begin{center}
\epsfig{file=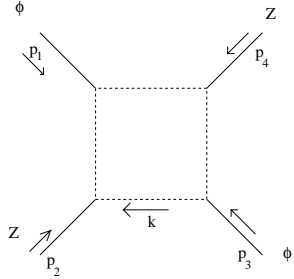,width=3.8cm}
\caption{Fermion loop contribution to the 4-point 
scalar vertex.\label{4pfermi}}
\end{center}
\end{figure}
%\begin{flushleft}
%\epsffile{kvadrat.eps}
%\end{flushleft}

$\bullet$ Figure \ref{4pfermi} with amplitude:
\begin{equation}
I_{square}=4\int d^4k { [k,k+p_2,k+p_1+p_2,k-p_3]\over
k^2(k+p_2)^2(k+p_1+p_2)^2(k-p_3)^2}
\end{equation}

\begin{figure}[ht]
\begin{center}
\epsfig{file=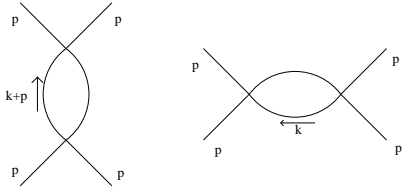,width=5.3cm}
\caption{Pure scalar contribution to the 4-point
scalar 
vertex.\label{fish}}
\end{center}
\end{figure}
%\begin{flushleft}
%\epsffile{ryba.eps}
%\end{flushleft}

$\bullet$ Figure \ref{fish} with amplitude:
\begin{equation}
I_{fish}=-\int d^4k {
k^2(k+p_1+p_2)^2+(k-p_3)^2(k+p_2)^2\over
k^2(k+p_2)^2(k+p_1+p_2)^2(k-p_3)^2}
\end{equation}

\begin{figure}[ht]
\begin{center}
\epsfig{file=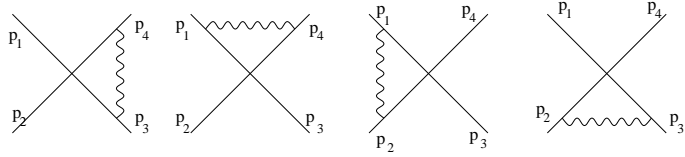,width=9cm}
\caption{Scalar and vector contribution to the 4-point
scalar 
vertex.\label{triangle}}
\end{center}
\end{figure}
%\begin{flushleft}
%\epsffile{triangle.eps}
%\end{flushleft}

$\bullet$ Figure \ref{triangle} with amplitude:
\begin{eqnarray}
I_{triangle}=&&{1\over 2}
\int d^4k {1\over k^2(k+p_2)^2(k+p_1+p_2)^2(k-p_3)^2}
\times\nonumber\\[5pt]&&\times
\left[(k+p_2)^2((k+p_3)\cdot (k+p_3+2p_1+2p_2))+
\right.\\[5pt]&& \left.
+k^2((k+p_2-p_1)\cdot(k-p_1-p_2-2p_3))+
\right.\nonumber\\[5pt]&& \left.
+(k-p_3)^2((k-p_2)\cdot (k+2p_1+p_2))+
\right.\nonumber\\[5pt]&& \left.
+(k+p_1+p_2)^2((k+2p_2)\cdot (k-2p_3))\right]
\end{eqnarray}

Since we are interested in finding the $Z$-factor of
the full graph,
we will discard terms that are finite after this
vertex correction is 
inserted
into the larger diagram. In particular, using power
counting, it is 
easy to see that
we will need to keep terms proportional to $k^4$,
$k^3p_3$ and 
$p_3^2$. The last two terms are divergent only after
the one loop graph
is inserted in the two-loop one, while the first term
leads to 
divergences 
already at the one-loop level.  Using these
observations,
the sum of all these diagrams has the following
expression:
\begin{eqnarray}
&&I_{square}+I_{fish}+I_{triangle}=\nonumber\\[5pt]=
&&\int d^4k 
{4 k^4-4k^2(k\cdot p_3)+
p_3^2[(k\cdot p)-2(k\cdot p_2)-(p\cdot p_2)]+
\ldots\over
k^2(k+p_2)^2(k+p_1+p_2)^2(k-p_3)^2}
\end{eqnarray}
where the dots stand for the terms which are less than
quadratic in 
$p_3$ and 
therefore are convergent in the full graph. The
integral over $k$ is 
divergent 
because of the first term in the numerator. Thus,
finiteness of the 
vertex 
correction requires introduction of a counterterm 
equal to $-{4\over \epsilon}$.

The full diagram (figure \ref{2leg}) receives
contributions from two
one-loop counterterms. First, we have the counterterm
we have just 
introduced which brings
\begin{equation}
I_{ct,\lambda}=
-{4\over\epsilon}\left({1\over\epsilon}+2-C+\log{1\over
p^2}\right)~~.
\end{equation}
Furthermore, there is the contribution due to the
counterterm to the 
one-loop interaction exchanging the positions of
$\phi$ and $Z$. In the 
two-loop context it produces:
\begin{figure}[ht]
\begin{center}
\epsfig{file=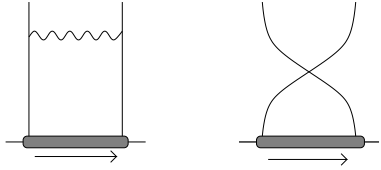,width=5cm}
\caption{Exchange counterterm contribution to the
``two interacting 
legs'' diagrams.}
\end{center}
\end{figure}
%\begin{flushleft}
%\epsffile{ct01.eps}
%\end{flushleft}
\begin{eqnarray}
&&I_{ct}={1\over 2\epsilon}
\int d^4k {(k+p_2)^2-(k-p_2)\cdot(k+p_2+2p_1)\over
k^2(k+p_2)^2(k+p)^2}
=\nonumber\\[5pt]&&=
{1\over\epsilon}
\int d^4k {(k\cdot p_2)-(k\cdot p_1)+(p\cdot p_2)\over
k^2(k+p_2)^2(k+p)^2}
=\\[5pt]&&=
\int d^4k d^4p_3{p_3^2[(k\cdot p_2)-(k\cdot
p_1)+(p\cdot p_2)]\over
p_3^2(k-p_3)^2(p_3+p)^2
k^2(k+p_2)^2(k+p)^2}
\end{eqnarray}
where the last equal sign holds only up to finite
terms.

\begin{figure}[ht]
\begin{center}
\epsfig{file=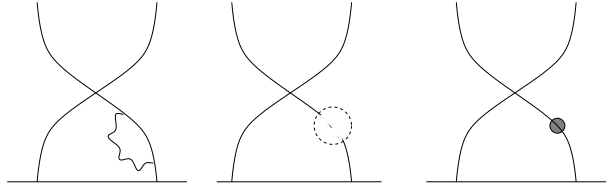,width=8cm}
\caption{Scalar self-energy contribution to the ``two
interacting 
legs'' diagrams
\label{selfenergy}}
\end{center}
\end{figure}

Putting everything together we find that
\begin{eqnarray}
I_{square}+I_{fish}+I_{triangle}+I_{ct}+I_{ct,\lambda}=
-{2\over\epsilon^2}+(finite)
\label{boxndstuff}
\end{eqnarray}
%We have:
%\begin{eqnarray}
%&&I_{square}+I_{fish}+I_{triangle}+I_{ct}=
%4\int dk dp_3{k^2-(k\cdot p_3)\over
%p_3^2(p+p_3)^2(k-p_3)^2(k+p_2)^2(k+p)^2}=\nonumber\\[5pt]&&=
%{2\over\epsilon^2}+{4\over\epsilon}\left(2-C+\log
%{1\over p^2}
%\right)
%\end{eqnarray}
This is not the whole contribution of diagrams with
two interacting 
legs
since we have to take into account the correction to
the scalar 
propagator.
The half of the relevant diagrams, including their
counterterm, are 
pictured in 
figure \ref{selfenergy}.
%\begin{center}
%\epsffile{se.eps}
%\end{center}
The filled circle represents the counterterm due to
wave function 
renormalization. The other half 
have the self-energy and counterterm graphs inserted
on the left leg.
These diagrams contribute:
\begin{equation}
\label{se}
I_{se}=-4\int d^4k 
\left[\left({1\over\epsilon}+2-C\right){1\over
(k^2)^{1+\epsilon}}
-{1\over\epsilon}{1\over k^2}\right]{1\over(k+p)^2}=
{2\over\epsilon^2}-{2\over\epsilon}+(finite)
\end{equation}

Combining equations (\ref{boxndstuff}) and (\ref{se}),
we find that the divergent part of the diagrams with
only two 
interacting legs is:
\begin{equation}
I_{2-legs}=I_{square}+
I_{fish}+I_{triangle}+I_{ct}+I_{ct,\lambda}+I_{se}=
-{2\over\epsilon}+\mbox{finite}
\end{equation}

The remaining diagrams are those which involve three
fields. 
We have already computed 
all the necessary integrals except for the graphs on
Fig. 16.
\begin{figure}[ht]
\begin{center}
\epsfig{file=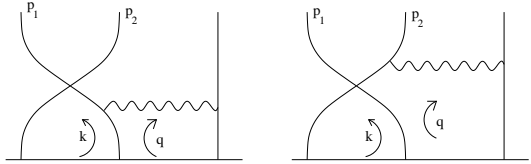,width=7cm}
\caption{Diagrams with vector boson exchange.
\label{vv'}}
\end{center}
\end{figure}
%\begin{flushleft}
%\epsffile{morediag.eps}
%\end{flushleft}
The left graph has the following amplitude:
\begin{eqnarray}
&&I_v={1\over 2}\int d^4q d^4k 
{(q\cdot(q+2k))\over k^2(k+p_1+p_2)^2(k+q)^2 q^4}
=\nonumber\\[5pt]&&=
{1\over 2}\int d^4k {1\over k^2(k+p_1+p_2)^2}\left[
{\Gamma(\epsilon)\Gamma(1-\epsilon)^2\over\Gamma(2-2\epsilon)}
{1\over [k^2]^{\epsilon}}-
2{\Gamma(1+\epsilon)\Gamma(1-\epsilon)^2\over\Gamma(2-2\epsilon)}
{1\over [k^2]^{\epsilon}}\right]+\ldots
=\nonumber\\[5pt]&&=
{1\over 2}
{\Gamma(\epsilon)\Gamma(2\epsilon)\Gamma(1-\epsilon)^3\over
\Gamma(1+\epsilon)\Gamma(2-3\epsilon)}
{1\over [(p_1+p_2)^2]^{2\epsilon}}+\ldots
=\nonumber\\[5pt]&&=
{1\over 2}\left[{1\over 2\epsilon^2}-{1\over
2\epsilon}+
{1\over\epsilon}
\left(2-C+\log{1\over
(p_1+p_2)^2}\right)\right]+\ldots
\end{eqnarray}
where dots stand for the finite part. The amplitude
for the right diagram 
is:
\begin{eqnarray}
&&I_v'={1\over 2}\int d^4k d^4q {(q\cdot(q-2p_2))\over
(k+p_1+p_2)^2(k+q)^2(q-p_2)^2 q^4}
=\nonumber\\[5pt]&&={1\over 2}
{\Gamma(\epsilon)\Gamma(1-\epsilon)^2\over\Gamma(2-2\epsilon)}
\int d^4q {(q\cdot(q-2p_2))\over
[(q-p_1-p_2)^2]^{\epsilon}
(q-p_2)^2q^4}
=\nonumber\\[5pt]&&=
{1\over 2}\left[{1\over 2\epsilon^2}-{3\over
2\epsilon}
+{1\over\epsilon}\left(2-C+\log{1\over
p_2^2}\right)\right]+\ldots
\end{eqnarray}
Besides the diagrams with exchange of gauge field,
there are also those 
which 
involve only scalar couplings. Their amplitude is, up
to numerical 
factors, the same as the amplitude of the graph in
figure \ref{e2iphi}.
For this reason we will introduce the notation $I_s$
for the scalar 
diagram:
\begin{equation}
I_s={1\over
4\epsilon^2}{\Gamma(1+2\epsilon)\Gamma(1-\epsilon)^3\over
(1-3\epsilon)(1-2\epsilon)\Gamma(1-3\epsilon)}=
{1\over 2}\left[{1\over 2\epsilon^2}+{1\over
2\epsilon}
+{1\over\epsilon}(2-C)\right]
\end{equation}
and express everything in terms of $I_v$, $I_{v'}$ and
$I_s$. 
To summarize, we have the following building blocks:
\begin{eqnarray}
&&I_s={1\over 2}\left[{1\over 2\epsilon^2}+{1\over
2\epsilon}
+{1\over\epsilon}\left(2-C+\log{1\over p^2}\right)
\right]\nonumber\\[5pt]
&&I_v={1\over 2}\left[{1\over 2\epsilon^2}-{1\over
2\epsilon}+
{1\over\epsilon}\left(2-C+\log{1\over(p_1+p_2)^2}\right)\right]
\nonumber\\[5pt]
&&I_{v'}={1\over 2}\left[{1\over 2\epsilon^2}-{3\over
2\epsilon}
+{1\over\epsilon}\left(2-C+\log{1\over
p_2^2}\right)\right]\nonumber
\end{eqnarray}
It is convenient to organize the remaining diagrams
with their
counterterms in four groups. Each of them leads at
most
to local divergencies. 

\begin{figure}[ht]
\begin{center}
\epsfig{file=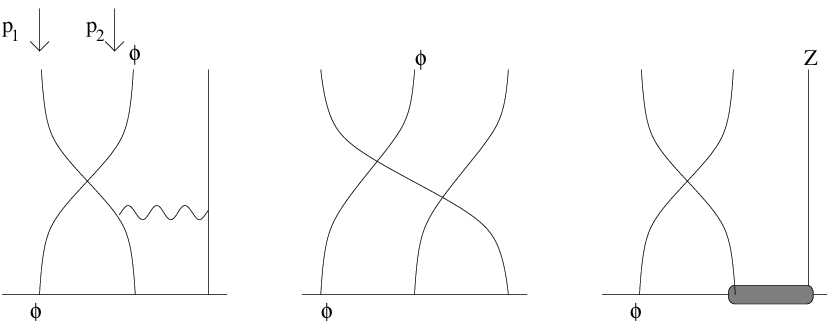,width=7.5cm}
\caption{Group 1. \label{g1}}
\end{center}
\end{figure}
%\begin{center}
%\epsffile{group1.eps}
%\end{center}
$\bullet$ Group 1 (figure \ref{g1}) with amplitude
\begin{equation}
I_v+I_s-{1\over\epsilon}\left({1\over\epsilon}+
2-C+\log{1\over (p_1+p_2)^2}\right)=-{1\over
2\epsilon^2}+(finite)
\end{equation}

\begin{figure}[ht]
\begin{center}
\epsfig{file=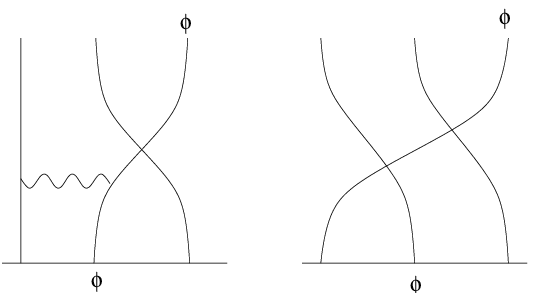,width=5cm}
\caption{Group 2. \label{g2}}
\end{center}
\end{figure}
%\begin{center}
%\epsffile{group2.eps}
%\end{center}
$\bullet$ Group 2 (figure \ref{g2}) with amplitude:
\begin{equation}
I_v-I_s=-{1\over 2\epsilon}+(finite)
\end{equation}

\begin{figure}[ht]
\begin{center}
\epsfig{file=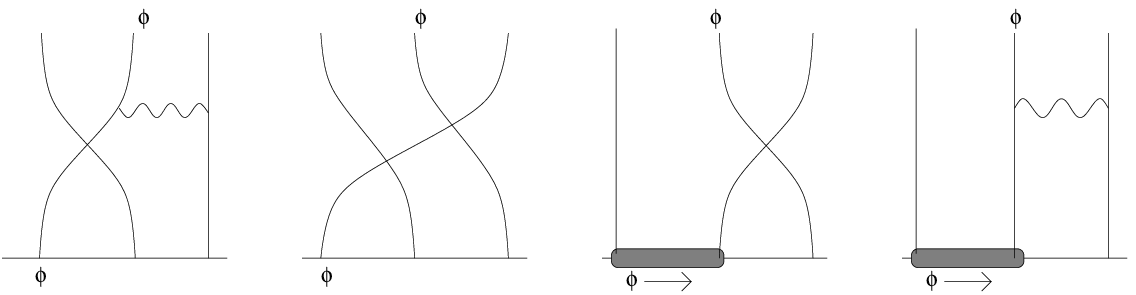,width=10cm}
\caption{Group 3. \label{g3}}
\end{center}
\end{figure}
%\begin{flushleft}
%\epsffile{group3.eps}
%\end{flushleft}
$\bullet$ Group 3 (figure \ref{g3}) with amplitude:
\begin{equation}
I'_v-I_s+{1\over 2\epsilon}\left({1\over\epsilon}+2-C+
\log{1\over p_2^2}\right)
-{1\over
2\epsilon}\left({1\over\epsilon}-C+\log{1\over
p_2^2}\right)
=0+(finite)
\end{equation}

\begin{figure}[ht]
\begin{center}
\epsfig{file=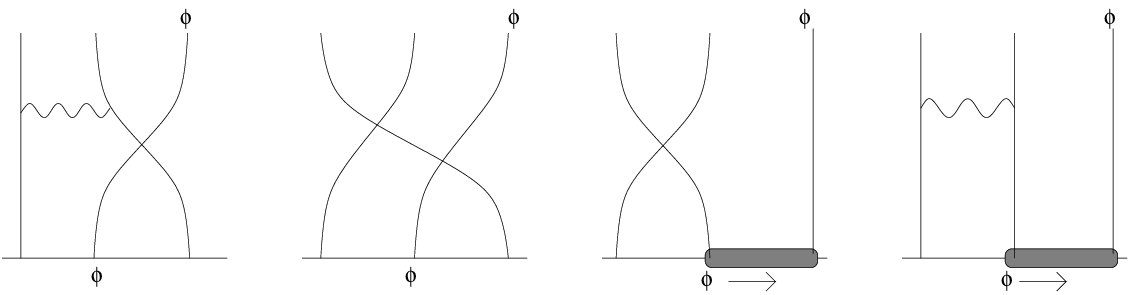,width=10cm}
\caption{Group 4. \label{g4}}
\end{center}
\end{figure}
%\begin{flushleft}
%\epsffile{group4.eps}
%\end{flushleft}
$\bullet$ Group 4 (figure \ref{g4}) with amplitude:
\begin{equation}
I'_v+I_s
-{1\over
2\epsilon}\left({1\over\epsilon}+2-C+\log{1\over
p_1^2}\right)
-{1\over
2\epsilon}\left({1\over\epsilon}-C+\log{1\over
p_1^2}\right)=
-{1\over 2\epsilon^2}+{1\over 2\epsilon}+(finite)
\end{equation}

Adding the contribution of the four groups we find
that their amplitude 
is:
\begin{equation}
I_{3-legs}=-{1\over\epsilon^2}+\mbox{finite}
\end{equation}

\begin{figure}[ht]
\begin{center}
\epsfig{file=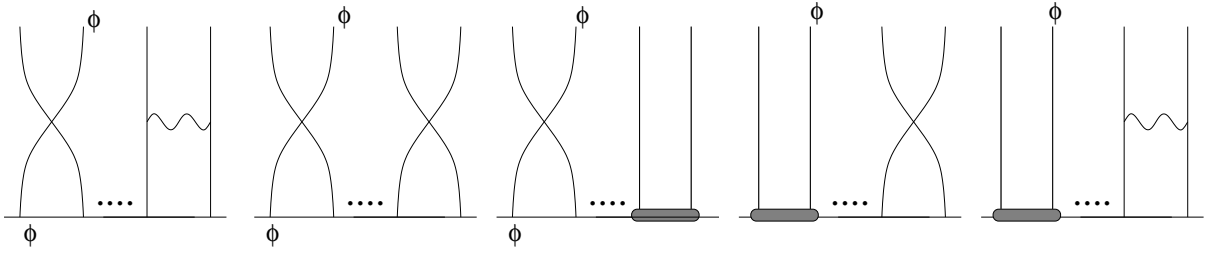,width=16cm}
\caption{Diagrams containing disconnected one-loop 
subdiagrams. \label{nonlocal}}
\end{center}
\end{figure}

The third and last set of diagrams contains
disconnected one-loop 
subdiagrams. The relevant ones are depicted in figure
\ref{nonlocal}.
%\begin{center}
%\epsffile{separated.eps}
%\end{center}
The divergent part of their amplitude is:
\begin{equation}
I_{non-local}=-{1\over\epsilon^2}(J-3)
\end{equation}
Adding up all the two loop diagrams, we find the
following divergent part:
\begin{equation}\label{ITwoLoopOne}
I_{2-loop,\;e^{i\varphi}}=I_{2-legs}+I_{3-legs}+I_{non-local}=
-{1\over\epsilon^2}(J-2)-{2\over\epsilon}
\end{equation}
from where the $Z$-factor proportional to
$e^{i\varphi}$ can be 
extracted.

\subsection{Renormalization of ${\cal O}$ at two
loops.}

We now turn to the renormalization of the operator
${\cal O}$ at two 
loops.
Adding (\ref{ITwoSubtr}) and (\ref{ITwoLoopOne}) we
find the divergent
part of the two loop diagrams with the insertion of
${\cal O}$:
\begin{equation}
\lambda^2\left[
e^{2i\varphi}
\left(-{1\over 2\epsilon^2}+{1\over 2\epsilon}\right)+
e^{i\varphi}\left(-{J-2\over\epsilon^2}-{2\over\epsilon}\right)
+{A_{2,2}\over\epsilon^2}+{A_{2,1}\over\epsilon}
+\mbox{finite}\right]
\end{equation}
Here ${A_{2,2}\over\epsilon^2}+{A_{2,1}\over\epsilon}$
is the 
contribution of the diagrams without the exchange of 
$Z$ and $\phi$ which we have not computed explicitly.

From here and equation (\ref{ZO})
we find that $Z_{\cal O}$ in equation (\ref{renop}) 
has the following two-loop expression:
\begin{eqnarray}\label{ZOTwoLoop}
&&Z_{{\cal O},\;2-loop}=
1-{\lambda\over\epsilon}(e^{i\varphi}+e^{-i\varphi}
+(J-2))+\nonumber\\[5pt]&&
+{\lambda^2\over\epsilon^2}\left({1\over
2}(e^{2i\varphi}+e^{-2i\varphi})
+(J-2)(e^{i\varphi}+e^{-i\varphi})-A_{2,2}\right)
+\\[5pt]&&
+{\lambda^2\over\epsilon}\left(-{1\over
2}(e^{2i\varphi}+e^{-2i\varphi})
+2(e^{i\varphi}+e^{-i\varphi})-A_{2,1}\right)
\end{eqnarray}
We are left with the task of determining the various
unknown 
coefficients in the above equation. One way of
determining them 
is doing the explicit computation and finding the
contribution of 
diagrams with no exchange of $Z$ and $\phi$. This is a
rather 
tedious exercise, since the number of diagrams is
substantially 
larger than those considered by now. The easier way is
to notice that,
%Taking into account the result (\ref{ORenOneLoop}) we
%have
%the following two-loop renormalization of ${\cal O}$:
%\begin{equation}
%{\cal O}=Z_{{\cal O},\;2-loop}{\cal O}^{ren}
%\end{equation}
%with
%\begin{eqnarray}\label{ZOTwoLoop}
%&&Z_{{\cal O},\;2-loop}=
%1-{\lambda\over\epsilon}(e^{i\varphi}+e^{-i\varphi}
%+(J-2))+\nonumber\\[5pt]&&
%+{\lambda^2\over\epsilon^2}\left({1\over
%2}e^{2i\varphi}
%+(J-2)e^{i\varphi}-A_{2,2}\right)
%+{\lambda^2\over\epsilon}\left(-{1\over
%2}e^{2i\varphi}
%+2e^{i\varphi}-A_{2,1}\right)
%\end{eqnarray}
for $\varphi=0$ our operator is BPS and therefore
has vanishing anomalous dimension. This means that the
$Z$ factor for
$\varphi=0$ cancel against the renormalization of
external 
lines. In other words
\begin{equation}
\left.Z_{\phi}^{-J/2}Z_{\cal
O}\right|_{\varphi=0}=1~~.
\end{equation}
Combining this with equation (\ref{ZOTwoLoop}) we get 
\begin{equation}
Z_{\phi}^{-J/2}Z_{\cal O}=
\exp\left\{{1\over\epsilon}
\left[-\lambda(e^{i\varphi}+e^{-i\varphi}-2)
-{\lambda^2\over 2}(e^{i\varphi}+e^{-i\varphi}-2)^2+
\ldots\right]\right\}
\end{equation}
which exactly agrees with equation (\ref{Conjecture}).

To summarize this section, 
we have verified at the two loop level that
the renormalization of ${\cal O}$ depends on the
coupling constant
$\lambda$ only in the combination 
$\lambda(e^{i\varphi}+e^{-i\varphi}-2)$.

\section{Higher orders}

In the previous section we have computed the two-loop
contribution to 
the 
anomalous dimension of operators dual to stringy
excitations and found 
that the conjecture put forward in section
\ref{generalandim} holds at 
this
order. The contribution of the many diagrams leading
to this result 
is quite entangled and a pattern of cancellations that
can be 
generalized 
to all loop order does not seem to emerge. However,
{\it assuming} 
that the conjecture (\ref{Conjecture}) holds, we can
actually derive
the anomalous dimension to all orders in perturbation
theory.
The idea is that, given  this assumption, there is
exactly one relevant
diagram per loop order. In this section we will
compute all these 
diagrams.
We will  sum these graphs and argue that the anomalous
dimension is indeed given by (\ref{YangMills}).

\subsection{Subset of diagrams.}

\begin{figure}[ht]
\begin{center}
\epsfig{file=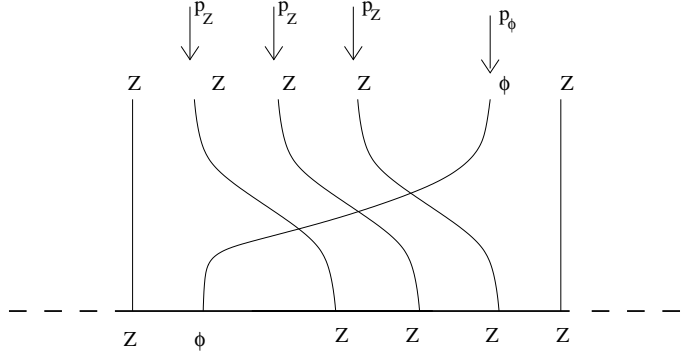,width=9cm}
\caption{The $j$-loop diagram contribution 
to $c_{j,j}$, $j=3$ is shown.\label{arbitorder}}
\end{center}
\end{figure}

By briefly studying the diagrams at an arbitrary loop
order $L$, 
it is not difficult to see that there is only one that
requires 
a counterterm proportional to $e^{iL\varphi}$. This
diagram
involves only scalar field couplings and at each
vertex $\phi$ and $Z$ 
switch places. 
A generic graph of this type is shown in figure
\ref{arbitorder}. 
Under the assumption stated above, this diagram gives
the only 
contribution to the anomalous dimension at L loops.

We will now compute the contribution of figure
\ref{arbitorder}.
In principle one should put arbitrary momenta on
external lines.
%\begin{center}
%\leavevmode
%\epsffile{threeloop.eps}
%\end{center}
However, as long as infrared divergences are not
encountered, we 
can make further simplifying assumptions, in
particular, we can set 
some momenta to zero since this does not change the
divergent part. 
In figure \ref{arbitorder} we put different momenta on
the $Z$ legs 
and on the $\phi$ legs. Using the argument above we
will compute 
this diagram in the limit $p_{Z}\to 0$.
The corresponding Feynman integral is:
\begin{equation}\label{MultiIntegral}
I_n(p_{\phi})=
\int {d^{4-2\epsilon} r_1\over r_1^2(r_1-r_2)^2}
\int {d^{4-2\epsilon} r_2\over r_2^2(r_2-r_3)^2}
\cdots
\int {d^{4-2\epsilon} r_n\over r_n^2(r_n-p_{\phi})^2}
\end{equation}
Probably the easiest way of computing this integral is
to set up
a recurrence relation based on the following identity:
\begin{equation}
\int d^4q {1\over [q^2]^{1+n\epsilon}(q-p)^2}=
{\Gamma(1-\epsilon)\Gamma((n+1)\epsilon)\Gamma(1-(n+1)\epsilon)\over
\Gamma(1+n\epsilon)\Gamma(2-(n+2)\epsilon)}{1\over 
[p^2]^{(n+1)\epsilon}}
\end{equation}
Then, solving the recurrence, we find that the
integral 
(\ref{MultiIntegral}) 
is given by:
\begin{equation}\label{MultiLoop}
I_n(p_{\phi})={1\over
(n-1)!}{\Gamma(n\epsilon)\over\epsilon^{n-1}}
{\Gamma(1-\epsilon)^{n+1}\over
\Gamma(2-(n+1)\epsilon)}
\times{1\over (1-2\epsilon)\cdots(1-n\epsilon)}\times
{1\over [p_{\phi}^2]^{n\epsilon}}
\end{equation}
We now have all ingredients that will lead to the
proof that
equation (\ref{YangMills}) holds to all orders in
perturbation theory

\subsection{The renormalized sum of the diagrams.}

Assuming that the coupling constant enters the formula for the
anomalous dimension in the combination 
${\hat\lambda}=\lambda (e^{i\varphi}+e^{-i\varphi}-2)$ as stated in 
equation (\ref{Conjecture}), 
the $Z$-factor for an operator ${\cal O}$ of anomalous 
dimension $c({\hat\lambda})$:
\begin{equation}
Z=\exp\left[{1\over\epsilon}\int\limits_0^1
{dt\over 2t}\left(c(t{\hat\lambda})-1\right)
\right]
\end{equation}
Showing that $c-1$ is indeed its anomalous dimension amounts to showing 
that the product between $Z$ and the sum of regularized Feynman integrals 
corresponding to diagrams with one insertion of some operator ${\cal O}$
is finite. Under our assumption about the dependence on $\lambda$ 
the sum of regularized Feynman integrals is:
\begin{equation}
I(p)=1+\sum\limits_{n=1}^{\infty} 
{1\over n!}{ {\hat\lambda}^n\over\epsilon^n} 
{\Gamma(1+n\epsilon)\Gamma(1-\epsilon)^{n+1}\over
\Gamma(2-(n+1)\epsilon)
\prod\limits_{k=2}^n(1-k\epsilon)}
{1\over [p_{\phi}^2]^{n\epsilon}}
\end{equation}
We want to show that 
\begin{equation}
\sqrt{1-4\lambda(e^{i\varphi}+e^{-i\varphi}-2)}-1
\end{equation}
is the anomalous dimension of ${\cal O}$. 
This is equivalent to the statement that 
\begin{equation}\label{RenSum}
I^{ren}(p)=\exp\left[{1\over\epsilon}\int\limits_0^1
{dt\over 2t}(\sqrt{1-4t{\hat \lambda}}-1)\right]
I(p)
\end{equation}
is finite as $\epsilon\rightarrow 0$. 
One can indeed verify order by order in $\hat\lambda$ that (\ref{RenSum})
is finite. There is also a general argument which we now describe.

Notice that the integral in the exponent can be 
explicitly computed and it gives
\begin{equation}\label{Z}
\exp\left[{1\over\epsilon}
\int_0^1 {dt\over 2t} (\sqrt{1-4{\hat\lambda} t}-1)\right]=
\exp\left\{{1\over\epsilon}
\left[\sqrt{1-4{\hat\lambda}}-1+\log{2\over 1+\sqrt{1-4{\hat\lambda}}}
\right]\right\}
\end{equation}
This has to be canceled by a similar contribution coming from the 
sum of Feynman diagrams.
To find the leading exponential behaviour of $I(p)$ we represent 
the sum as an integral and use the saddle point approximation:
\begin{equation}
\begin{array}{l}
I(p)\sim
\int\limits^{\infty}{1\over\epsilon}dx
{\Gamma(1+x)\over\Gamma(2-x)}
{\left({{\hat\lambda}\over\epsilon}\right)^{x\over\epsilon}\over
(x/\epsilon)!}
\exp\left[-{1\over\epsilon}\left((1-x)\log(1-x)+x)\right)\right]
=\\[5pt]=
{1\over\epsilon}\int\limits^{\infty}dx\sqrt{\epsilon\over x}
f(x)
\exp\left[{1\over\epsilon}\left(
x(\log({\hat\lambda})-\log x) + 2x + (1-x)\log(1-x)\right)\right]
\end{array}
\end{equation}
Here we have defined $x=n\epsilon$, and $f(x)$ is finite at $\epsilon\to 0$.
The equation for the saddle point is then:
\begin{equation}
\log x_0(1-x_0)=\log {\hat\lambda}
\end{equation}
This gives
$x_0={1-\sqrt{1-4{\hat\lambda}}\over 2}$. The series then is approximated
by the saddle point:
\begin{equation}
I(p)\sim
\left[{1+\sqrt{1-4{\hat\lambda}}\over 2}e^{1-\sqrt{1-4{\hat\lambda}}}
\right]^{1/\epsilon}~~.
\end{equation}
which  is exactly the inverse of (\ref{Z})! 
The pre-exponential factor correcting the saddle point approximation 
is a series in $\epsilon$, and it is finite at $\epsilon\to 0$. 
This proves that (\ref{RenSum}) is finite at $\epsilon=0$.

\section{Conclusions.}
The usual 't~Hooft perturbation theory for large $N$
field theory
is an  expansion in the small parameter
$\lambda=g^2N$.
Our results indicate that there are amplitudes
for which the actual small parameter is not $\lambda$
but rather $\lambda\sinh^2\left({\varphi\over
2}\right)$. 
In these amplitudes, we can consider the regime when
$\lambda$ is large 
and $\varphi$ is small, so that
$\lambda\varphi^2\ll 1$. In this
regime, the perturbative expansion is still valid even
though
the 't~Hooft parameter is large. On the other hand, a
large 'tHooft
parameter implies that the calculation can be done in
the dual
picture, namely  string theory in  a plane wave
background. 
This means that the perturbative computation in the
field theory  
can be compared with the perturbative computation on
the string 
worldsheet.

We have partialy summed the perturbation series and
found that
  the result for the anomalous  dimension is indeed
equal
to the mass of the string excitation. This is true
order
by order in the Yang-Mills perturbation theory,
presumably up
to  order roughly equal to $J$ (the R charge of the
operator). 
At that   order, $\simeq J$, we should include 
diagrams involving
the collision of two $\phi$'s. It is possible that in
these diagrams,
the assumption that $\lambda$ enters in the
combination
$\lambda\sinh^2\left({\varphi\over 2}\right)$ breaks
down and therefore
one cannot rely on   perturbation theory.

Given this success of the BMN proposal it would be
very interesting to extend the
perturbative calculations of the anomalous dimensions
of these operators,
in the large $N$ and $J$ limit, to include non-planar
corrections of order $g^2$.,
These should correspond on the string side to string
loop corrections to the
masses of string excitations. In this way one could
\lq \lq derive\rq\rq
 the interacting string theory in a plane wave
background from   gauge theory.

\section{Acknowledgements.}
We want to thank Tibor Kucs for discussions about two
loop momentum
integrals. 
The work of A.M. and D.G. was supported in part by the
National
Science Foundation under Grant No. PHY99-07949.
The work of A.M. was partly supported by the RFBR
Grant No. 
00-02-116477
and in part by the Russian Grant for the support of
the
scientific schools No. 00-15-96557.
The work of RR was supported in part by DOE under
Grant No.
91ER40618(3N) and in part by the National Science
Foundation 
under Grant No. PHY00-9809(6T).

\end{document}